\documentclass[11pt, a4paper]{article}
\usepackage{jheppub}
\usepackage{amsmath,amssymb}
\usepackage{latexsym}
\usepackage{multicol}
\usepackage{graphicx}
\usepackage{hyperref}
\usepackage{caption}
\usepackage{bm}
\allowdisplaybreaks[1]
\newcommand{\R}{\mathbb{R}}

\newcommand{\del}{\partial}
\newcommand{\be}{\begin{equation}}
\newcommand{\ee}{\end{equation}}

\newcommand{\half}{\frac{1}{2}}
\newcommand{\nn}{\nonumber}
\newcommand\fr[1]{\frac{1}{#1}}
\usepackage{array}
\usepackage{tikz}
\usetikzlibrary{trees}
\usepackage{siunitx}
\usepackage{varwidth}
\usepackage{pifont}
\usepackage{cancel}
\usetikzlibrary{shapes.geometric, arrows, shadows}

\tikzstyle{forces} = [rectangle, rounded corners, minimum width=2cm, minimum height=0.8cm,text centered, draw=black, fill=white]
\tikzstyle{spin} = [rectangle, rounded corners, minimum width=1.3cm, minimum height=0.8cm,text centered, draw=black, fill=white]
\tikzstyle{theory} = [ellipse, minimum width=2.4cm, minimum height=0.8cm,text centered, draw=black, fill=white]
\tikzstyle{arrow} = [thick,->,>=stealth]
\tikzstyle{arrow1} = [thick,<->,>=stealth]
\tikzstyle{line} = [draw, -latex']
\tikzset{
	double arrow/.style args={#1 colored by #2 and #3}{
		-stealth,line width=#1,#2, 
		postaction={draw,-stealth,#3,line width=(#1/2),
			shorten <=(#1)/3,shorten >=2*(#1)/3}, 
	}
}

\newcommand{\ndt}{\noindent}
\newcommand{\delm}{{\partial_-}}

\def\bea{\begin{eqnarray}}
\def\eea{\end{eqnarray}}
\begin{document}
	\vskip 1cm
	\title{\Large{\bf{Ehlers symmetry in four dimensions}}}
	\vskip 1.5cm
	\author{Sucheta Majumdar}
	\vskip 1.5cm
	\affiliation{Universit\'e Libre de Bruxelles and International Solvay Institutes, \\Campus Plaine $-$CP 231, B-1050 Bruxelles, Belgium}
	\vskip 1.5cm
	\emailAdd{sucheta.majumdar@ulb.ac.be}

	\abstract{\ndt Starting with the light-cone Hamiltonian for gravity, we perform a field redefinition that reveals a hidden symmetry in four dimensions, namely the Ehlers $SL(2,\R)$ symmetry. The field redefinition, which is non-local in space but local in time, acts as a canonical transformation in the Hamiltonian formulation keeping the Poisson bracket relations unaltered. We discuss the electro-magnetic duality symmetry of gravity in the light-cone formalism, which forms the $SO(2)$ subgroup of the Ehlers symmetry. The helicity states in the original Hamiltonian are not in a representation of the enhanced symmetry group. In order to make the symmetry manifest, we make a change of variables in the path integral from the helicity states to new fields that transform linearly under the $SO(2)$ duality symmetry.}
	\vfill

\maketitle

\section{Introduction}

Theories of gravity and supergravity have been known to exhibit rich hidden symmetries upon dimensional reduction from their higher-dimensional parent theories. Some of the celebrated examples include the Ehlers symmetry and the infinite-dimensional Geroch group in Einstein's gravity~\cite{Ehlers,Geroch:1970nt} and the exceptional symmetries in maximal supergravity theories~\cite{Cremmer:1979up,Marcus:1983hb,Damour:2002cu}. These hidden symmetries are believed to appear when the large spacetime symmetry group in the parent theory splits into a smaller one in lower dimensions, along with some internal symmetries that can further be enhanced using electro-magnetic duality. 

\vskip 0.2cm

 In the recent years, some of these symmetries have been realized in the parent theory prior to dimensional reduction~\cite{Hohm:2013jma,Hohm:2013pua,Hohm:2013uia,Hohm:2014fxa}. This offers a radically different perspective that these enhanced symmetries might already exist in the higher-dimensional parent theories, but may not manifest themselves in a local and Lorentz-covariant formulation. One can choose to make either the spacetime symmetries or the internal symmetries manifest at the level of the action. The light-cone formalism, where Lorentz covariance is not manifest, serves as a suitable platform to look for these symmetries in the higher-dimensional theories. The light-cone approach has, in fact, proved to be fruitful for maximal supergravity theories, where we have found strong indications for an $E_{7(7)}$ symmetry in eleven dimensions~\cite{Ananth:2016abv} and an $E_{8(8)}$ symmetry in four dimensions~\cite{Ananth:2017nbi} up to a non-trivial order in the perturbation constant.

\vskip 0.2cm
 When applied to Einstein's gravity, this approach led to a similar result for the Ehlers symmetry in four dimensions~\cite{Ananth:2018hif}, which is originally present only in three dimensions. However, what remained to be understood was how this version of light-cone gravity with a hidden $SL(2, \R)$ symmetry could be related to the original light-cone gauge-fixed Einstein-Hilbert action in four dimensions~\cite{Scherk:1974zm,Bengtsson:1984rv} . In this paper, we attempt to bridge this gap with the help of a field redefinition, which can be interpreted as a canonical transformation in the Hamiltonian formulation. The field redefinition involves some new operators that are non-local in space but local in time. The non-locality occurs only in the field redefinition and not at the level of the action. At a given order in coupling constant, the non-local terms so generated can be systematically eliminated by adding correction terms to the field redefinition. We show this explicitly to the second order in the coupling constant. 
\vskip 0.2cm
It is worthwhile to pause and contemplate why the issue of finding the correct field redefinition in the light-cone formalism is interesting. Although we set out to address a very specific problem in light-cone gravity, our motivation is to understand the origin of hidden symmetries in gravity and supergravity theories from a more fundamental standpoint. With the simple example of Ehlers symmetry, we propose a method that takes a given theory to a different formulation, where one can see a symmetry enhancement {\it without} any dimensional reduction and subsequent ``oxidation" back to four dimensions. The central piece of the puzzle is the $SO(2)$ subgroup of the $SL(2, \R)$, that represents the duality symmetry in gravity. The original light-cone action for gravity in four dimensions is expressed in terms of the two helicity states of the graviton, which are not suitable for representing the Ehlers symmetry. We must perform a field redefinition that maps these helicity states to new fields that transform linearly under the SO(2) duality group. Thus, our key result is that one must abandon the {\it helicity-invariant} configuration in favor of a {\it duality-invariant} one, in order to make the Ehlers symmetry manifest in four dimensions.

\vskip 0.2cm
 The paper is organized as follows. In section 2 we briefly review the work done in~\cite{Ananth:2018hif}, where the Ehlers symmetry was realized in light-cone gravity in four dimensions using the tools of dimensional reduction. In section 3 we present the non-local field redefinition that directly takes light-cone gravity in four dimensions to the $SL(2, \R)$ invariant version. We then argue that the field redefinition is a canonical transformation in the phase space with the correct Poisson brackets. We further show that the field redefinition merely amounts to a change of variables in the path integral. In section 4 we describe the light-cone representation of the Ehlers symmetry algebra and prove the invariance of the Hamiltonian under this symmetry. In section 5 we discuss the duality symmetry in light-cone gravity and explain how its action differs from the little group transformations in four dimensions. We conclude with a few remarks about the relevance of our results to the more sophisticated frameworks, such as the prepotential formalism, exceptional field theories etc., which explore these symmetry structures at greater lengths.

\section{Ehlers symmetry in light-cone gravity}
	
\ndt
We work with the Minkowski signature $\eta^{\mu \nu}= diag(-1, 1,1,1)$. The light-cone coordinates and derivatives are defined as
\bea
x^{\pm}~=~ \frac{x^0\, \pm\, x^3}{\sqrt 2} \ ,&& \del_{\pm}~=~ \frac{\del_0\, \pm\, \del_3}{\sqrt 2}	\ ,
\eea

\vskip 0.2cm
\ndt where $x^+$ is considered to be the time coordinate and $\del_+$ the time derivative. The transverse coordinates and derivatives are defined as	
\bea
x~=~ \frac{x^1\, +\, i\, x^2}{\sqrt 2}\ , && \bar x ~=~\frac{x^1\, -\, i\, x^2}{\sqrt 2} \nn
\eea
\ndt and 
\bea
\del~=~ \frac{\del_1\, +\, i\, \del_2}{\sqrt 2}\ , && \bar \del ~=~\frac{\del_1\, -\, i\, \del_2}{\sqrt 2} \nn
\eea
\ndt  such that
\be
\bar \del\, x~=~ \del \, \bar x~=~ 1\ .
\ee
\vskip 0.2cm
The light-cone gauge-fixing of Einstein-Hilbert action is presented in Appendix A. In a perturbative expansion in the coupling constant $\kappa$, the light-cone action for gravity in $d=4$ reads~\cite{Ananth:2007zy} 

\bea 
\label{d=4L}
\mathcal{S}&=&\int\, d^4\, x\, \Bigg\{\ \half \,\bar h\,\Box\, h\ + \, 2\,\kappa\, \bar{h}\, \del_-^2\left(-\,h\,\frac{\bar{\del}^2}{\del_-^2}h\,+\,\frac{\bar{\del}}{\del_-}h\,\frac{\bar{\del}}{\del_-}h\right) \, \nn \\
&&+\, 2\,\kappa\, h\, \del_-^2\left(-\,\bar h\,\frac{{\del}^2}{\del_-^2}\bar h\,+\,\frac{{\del}}{\del_-}\bar h\,\frac{{\del}}{\del_-}\bar h\right)\ +\ \mathcal O(\kappa^2)\, \Bigg\} \ ,
\end{eqnarray}
\vskip 0.2cm
\ndt where $h$ and $\bar h$ are the two helicity states of the graviton. Upon dimensional reduction, we can obtain the action for light-cone gravity in three dimensions. Gravity in three dimensions exhibits an $SL(2,R)$ symmetry due to Ehlers~\cite{Ehlers}, which acts like a non-linear sigma-model symmetry on the fields

\bea
\delta h=\frac{1}{\kappa}\,{\mbox {(constant)}} +\ \kappa\  \mbox{(quadratic in field)}\ +\ \cdots \ .
\eea 

\vskip 0.2cm
 The Ehlers symmetry can be regarded as a remnant of the $E_{8(8)}$ symmetry in the maximal supergravity theory in three dimensions. The $E_{8(8)}$ symmetry in $d=3$ can be decomposed into a linear $SO(16)$ part and a non-linear coset as
\bea
E_{8(8)}&=& SO(16)\ \times\ \frac{E_{8(8)}}{SO(16)}\ ,
\eea  

\ndt
which, upon supersymmetric truncation to pure gravity, yields
\bea
SL(2,\R)&=& SO(2)\ \times\ \frac{SL(2,\R)}{SO(2)}\ .
\eea

 However, the $SO(16)$ part of the $E_{8(8)}$ symmetry does not admit any interaction vertices of odd-order ($\kappa$, $\kappa^3$, $\ldots$) in the action . We, therefore, remove the three-point interaction terms obtained from the dimensional reduction of (\ref{d=4L}) through the following field redefinition
\bea \label{redef}
h\,\rightarrow\,h'-\kappa\,{\del_-^2}\,{\biggl (}\,\fr{\del_-}h'\,\fr{\del_-}h'\,{\biggr )}-2\kappa\,\fr{\del_-^2}\,{\biggl (}\,{\del_-^3}h'\,\fr{\del_-}{\bar h}'\,{\biggr )}\ +\ \mathcal{O}(\kappa^2) .
\eea

\vskip 0.2cm
\ndt
The field redefinition gives rise to some new interaction terms in the action involving time derivatives, which can be further eliminated by adding correction terms to 
(\ref{redef}). The action so obtained is invariant under the following set of transformations

\bea
L_+\, h'&=& \frac{1}{\kappa}\, a\ - \kappa\, a\,\frac{1}{\delm} (\delm h' \bar h') -2\,\kappa\,a\,\frac{1}{\del_-^2}({\del_-^3} h'  \frac{1}{\delm} \bar h'), \nn \\ [0.2cm] 
L_-\, h' &=&  -\, 2\, \kappa\ \bar a\ \frac{1}{\del_-^2} \left( \del_-^3 h' \, \frac{1}{\delm} \bar h' \right)\ +\ \fr{2}\ \kappa\ \bar a\  h'\ h'\ ,
\eea 

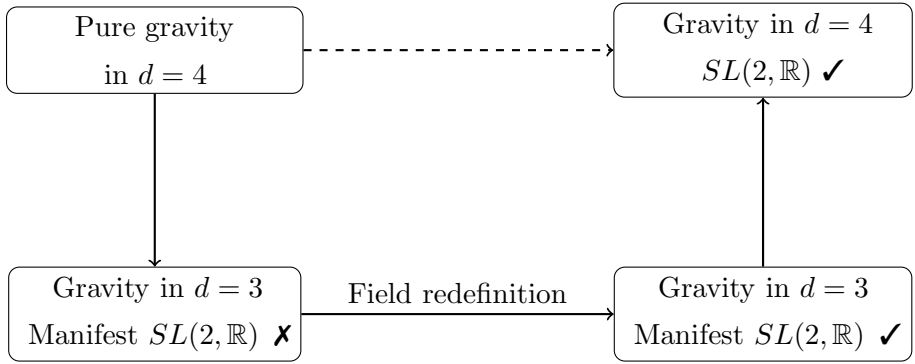
\begin{figure}[h!]
	\centering
	\begin{tikzpicture} 
	\node (gd4a) [forces] {\begin{varwidth}{14em} $\quad\quad$Pure gravity$ \qquad$\    \vskip 0.1cm $ \qquad \ \,\, $ in $d=4$ \qquad \ \end{varwidth}};
	\node (gd4b) [forces, xshift=8cm] {\begin{varwidth}{14em}$ \ \ \ $\ Gravity in $d=4$ $\ \  $  \vskip 0.1cm $ \ \qquad$  $ SL(2,\R)$ \ding{51} \quad  \end{varwidth}} ;
	\node (gd3a) [forces, below of = gd4a, yshift=-2.5cm] {\begin{varwidth}{14em}$\  $ \, Gravity in $d=3$$ \,  $  \vskip 0.1cm $  $ Manifest $SL(2,\R)\ \,  $\ding{55} \end{varwidth}};;
	\node (gd3b) [forces, xshift = 8cm, yshift=-3.5cm] {\begin{varwidth}{14em}$\ \  $  \,Gravity in $d=3 \quad $  \vskip 0.1cm $  $ Manifest $SL(2,\R)\, \  $\ding{51}   \end{varwidth}};
	\draw [->, line width=0.8pt , black] (gd4a) -- (gd3a);
	\draw [->, line width=0.8pt, black] (gd3b)-- (gd4b);
	\draw [->,line width=0.8pt, black] (gd3a) --  (gd3b) node[midway, above]{Field redefinition};
	\draw [dashed, ->,line width=0.8pt, black] (gd4a) --  (gd4b);
	\end{tikzpicture}
	\vskip 0.2cm
	\captionof{figure}{Ehlers symmetry in $d=4$ gravity}
	\label{fig:Ehlers}
\end{figure}

\vskip 0.2cm
\ndt
with the conjugate expressions for $\bar h'$. These transformations, which satisfy an $SL(2, \R)$ algebra, constitute the light-cone representation of the Ehlers symmetry in three dimensions. We can now invoke the quadratic form property of the Hamiltonian~\cite{Ananth:2017xpj} and ``oxidize" the theory to four dimensions keeping the Ehlers symmetry preserved in a way that is consistent with the Poincar\'e symmetry. We therefore arrive at a theory of light-cone gravity in $d=4$ with a hidden $SL(2, \R)$ symmetry. It appears as though there exist two distinct formulations of light-cone gravity in four dimensions with different symmetry structures, as depicted in figure \ref{fig:Ehlers}.
\vskip 0.5cm
\vskip 0.2cm

Our goal is to formulate a path from the original $d=4$ light-cone theory of gravity to the $SL(2, \R)$ invariant version, without having to go through the longer route involving dimensional reduction and ``oxidation". Such a path will connect the two different formulations of gravity in four dimensions and thereby serve as the missing link (dashed arrow) in the figure above. 
\section{Non-local field redefinition in $d=4$}

\ndt The field redefinition from light-cone gravity to the $SL(2, \R)$ invariant version involves some new non-local operators. Before we define these operators, let us consider the derivative structure of the light-cone Hamiltonian in $d=4$ 

\bea
\mathcal H& \sim & \del \bar h\, \bar \del h\ +\ \kappa  (\bar \del)^2 \, (h)^2\, \bar h\ +\ \kappa\ (\del)^2 \, (\bar h)^2\, h\ + \kappa^2 \, (\del\, \bar \del)\, (h)^2\, (\bar h)^2\ +\ \cdots
\eea 
\vskip 0.2cm

\ndt to demonstrate why we need to introduce them in the first place. In order to find a field redefinition that absorbs the cubic interaction vertices into the kinetic term, we have to define an operation which ``converts" a $\del$ into a $\bar \del$ and vice-versa. In the light-cone formalism, we use the $\frac{1}{\del_-}$ operator liberally to achieve the effect of removing a $\del_-$ derivative. We shall now define similar operators with the other two spatial derivatives.
\vskip 0.2cm
\subsection{Non-local operators}
\ndt In light-cone field theories, the $ \frac{1}{\del_-}$ operator is defined in terms of the Heaviside step function~\cite{Bengtsson:1984rv,Mandelstam:1982cb}. Consider two functions $g(x^-)$ and $f(x^-)$ such that

\bea
\del_-\, g(x^-)&=& f(x^-)\ .
\eea
\vskip 0.2cm
\ndt The $\frac{1}{\del_-}$ operator allows us to solve for $g(x^-)$ up to an arbitrary function $h$ independent of $x^-$

\bea 
g(x^-)~=~\frac{1}{\del_-}\, f(x^-)&=& \int^\infty_{-\infty} \, \epsilon (x^-- y^-) \, f(y^-)\, dy^- \ +\ h \ .
\eea

\vskip 0.2cm
The function $h$ can further be removed by imposing appropriate boundary conditions\footnote{We have imposed asymptotically Minkowski boundary conditions, i.e., $g^{\mu \nu}\rightarrow \eta^{\mu\nu}$ at infinity, which allow us to freely perform partial integration and drop the surface terms.}. This integral operator is non-local in the $x^-$ coordinate by construction. The function $f(x^-)$ can be recovered by acting with a $\del_-$ derivative on both sides of the above equation

\bea \label{op. defn.}
\del_- g(x^-)&=& \del_- \, \frac{1}{\del_-} f(x^-)~=~ f(x^-)\ . \\ && \nn
\eea
\vskip 0.2cm
Similarly, we can now extend this ``inverse of a derivative" operation to the other two spatial coordinates and define the operator

\bea {\label{del inv}}
\frac{1}{\del}\, f(\bar x)&=& \int^\infty_{-\infty} \, \epsilon (\bar x- \bar y) \, f(\bar y)\, d \bar y \, 
\eea
\vskip 0.2cm
\ndt and the conjugate operator

\bea
\label{del bar inv}
\frac{1}{\bar \del}\, f(x)&=& \int^\infty_{-\infty} \, \epsilon (x - y) \, f(y)\, dy \, ,
\eea

\vskip 0.2cm
\ndt that are local in time $x^+$ but non-local in the $\bar x$ and $x$ coordinates respectively.  In the momentum space one can define suitable pole prescriptions for $p, \bar p \rightarrow 0$ following~\cite{Mandelstam:1982cb}, which are relevant for loop calculations in the quantum theory\footnote{One encounters a similar kind of non-locality in the momentum space, while constructing MHV Lagrangians for Yang-Mills theory and gravity using field redefinitions~\cite{Ananth:2007zy, MHV Yang-Mills}.}. For the purposes of this paper, however, it suffices to treat these operators formally in the same fashion as in (\ref{op. defn.}). 

\vskip 0.2cm

\subsection{The field redefinition}
\ndt We work in a Hamiltonian framework starting with the light-cone Hamiltonian for gravity in $d=4$, which reads~\cite{Ananth:2007zy}

\bea \label{H4}
&\mathcal{H}&=~\del \bar h\, \bar \del h\ - \, 2\,\kappa\, \bar{h}\, \del_-^2\left(\,\frac{\bar{\del}}{\delm}h\,\frac{\bar{\del}}{\delm}h -\,h\,\frac{\bar{\del}^2}{\del_-^2}h\right) \,-\, 2\,\kappa\, h\, \del_-^2\left(\frac{{\del}}{\delm}\bar h\,\frac{{\del}}{\delm}\bar h -\,\bar h\,\frac{{\del}^2}{\del_-^2}\bar h\, \right)\ \nn \\[0.2cm]
&&- 2\, \kappa^2\, \Bigg\{ \frac{1}{\del_-^2} 
( \delm h \delm \bar h)\frac{\partial \bar \partial}{\del_-^2} 
( \delm h \delm \bar h)\ + \frac{1}{\del_-^3} ( \delm h \delm \bar h) \left( \partial \bar
\partial h \, \delm\bar h+ \delm h \partial \bar
\partial \bar h \right) \nonumber\\[0.2cm]
&&
-\frac{1}{{\delm}^2} 
( \delm h \delm \bar h )\, \left(2\, \partial \bar \partial h \, \bar h+ 2\, h \partial \bar
\partial \bar h + 9\, \bar \partial h \partial \bar  h + \partial h
\bar \partial \bar h   -  \frac{\partial \bar \partial}{\delm} h \, \delm\bar h
-  \delm h \frac{\partial \bar \partial}{\delm} \bar h 
\right)
\nonumber  \\ [0.2cm]
&&
-2 \frac{1}{\delm} (
2 \bar \partial h\, \delm \bar h 
+ h \delm \bar \partial \bar h 
- \delm \bar \partial h
\bar h ) 
\,h\,\partial \bar h   -2 \frac{1}{\delm} (
2 \delm h\, \partial \bar h
+ \delm  \partial h \, \bar h
- h \delm \partial \bar h ) \,\bar \partial h \, \bar h\nonumber 
\\ [0.2cm]
&&
-\frac{1}{\delm} (
2 \bar \partial h\, \delm \bar h 
+ h \delm \bar \partial \bar h 
- \delm \bar \partial h
\bar h )
\frac{1}{\delm} (
2 \delm h\, \partial \bar h
+ \delm  \partial h \, \bar h
- h \delm \partial \bar h ) \nonumber\\ [0.2cm]
&& - h\,\bar h\,\left(\partial \bar \partial h \, \bar  h + h
\partial \bar  
\partial \bar h + 2\, \bar \partial h \partial \bar  h 
+3 \frac{\partial \bar \partial}{\delm} h \, \delm\bar h
+3 \delm h \frac{\partial \bar \partial}{\delm} \bar h 
\right)\Bigg\}\ +\ \mathcal O(\kappa^3)\ .
\eea 
\vskip 0.2cm
\ndt In order to eliminate the three-point interaction vertices in the Hamiltonian, we perform the field redefinition

\bea \label{FRD}
h &=& C\ -\ \kappa \frac{\bar \del}{\del}\, \del_-^2\, \left(\frac{1}{\delm}C\, \frac{1}{\delm} C\right)\ -\ 2\, \kappa\, \frac{\del}{\bar \del}\, \frac{1}{\del_-^2}\, \left(\del_-^3 C\, \frac{1}{\delm} \bar C \right)\ ,\\
&& \nn
\eea

\vskip 0.2cm
\ndt which brings the Hamiltonian to the form

\bea 
\mathcal{H'}&=& \del \bar C\, \bar \del C\ +\ \mathcal{O}(\kappa^2)\  .
\eea

\vskip 0.2cm
The $\frac{1}{\del}$ and $\frac{1}{\bar \del}$ operators generate new terms in the Hamiltonian at order $\kappa^2$,  which are non-local in the $x$ and $\bar x$ coordinates. These terms can be removed by adding some $\kappa^2 $-correction terms to (\ref{FRD})

\bea\label{FRDcorr}
h &=& C\ -\ \kappa \frac{\bar \del}{\del}\, \del_-^2\, \left(\frac{1}{\delm}C\, \frac{1}{\delm} C\right)\ -\ 2\, \kappa\, \frac{\del}{\bar \del}\, \frac{1}{\del_-^2}\, \left(\del_-^3 C\, \frac{1}{\delm} \bar C \right)\ \nn  \\ [0.2cm]
&&-2\,\kappa^2\, \frac{1}{\bar \del \delm}\left[\frac{\del}{\delm}\bar C\, \frac{\bar \del}{\del}\,\del_-^4\,\left( \frac{1}{\delm}C \, \frac{1}{\delm}C\right)\right]	\ +\ 2\, \kappa^2\, \frac{\del}{\bar \del\del_-^2}\, \left[ C\, \frac{\bar \del}{\del}\, \del_-^4\, \left(\frac{1}{\delm}C \frac{1}{\delm}C\right)\right] \nn \\ [0.2cm]
&&+\ 4\, \kappa^2\, \frac{\del}{\bar \del \del_-^2}\, \left[\del_-^2 C\, \frac{\bar \del}{\del \del_-^2}\, \left(\del_-^3 \bar C\, \frac{1}{\delm}C\right)\right]\ +\ 4\,\kappa^2 \,\frac{\del}{\bar \del \del_-^2}\left[\bar C\, \frac{\del}{\bar \del}\, \left(\del_-^3 C\, \frac{1}{\delm}\bar C\right)\right]\ \nn \\[0.2cm]
&&-\ 4\, \kappa^2\, \frac{1}{\bar \del \delm}\, \left[\del_-^2 C\, \frac{\del^2}{\bar \del}\, \delm\, \left(\frac{1}{\delm}\bar C \, \frac{1}{\delm} \bar C\right)\right]\  +\ 4\, \kappa^2\, \frac{1}{\bar \del \delm}\, \left[\frac{\del^2}{\bar \del}(\bar C \del_-^2 C)\, \frac{1}{\delm} \bar C\right]  \nn \\ [0.2cm]
&&+\ 2\, \kappa^2\, \frac{\del}{\bar \del \del_-^2}\,\left[\del_-^2 C \, \frac{\del}{\bar \del}\, \del_-^2\, \left(\frac{1}{\delm}C\, \frac{1}{\delm}C\right)\right]\ .
\eea

\vskip 0.2cm
\ndt Thus, at any given order in the coupling constant, one can ensure that the Hamiltonian does not involve any non-local terms by correcting the field redefinition appropriately. The field redefinition also generates some local terms\footnote{By local terms, we mean the terms that do not have any non-locality other than the $\frac{1}{\delm}$-type and in particular, do not involve any $\frac{1}{\del}$ or $\frac{1}{\bar \del}$ operators.} which we add to the four-point vertex in the new Hamiltonian, which now reads

\bea \label{H4new}
\mathcal H'&=& \bar \del C\, \del \bar C\ + \mathcal{H}^{\kappa^2}\ + \ \kappa^2\, \del \del_-^2\, \left(\frac{1}{\delm}\bar C\, \frac{1}{\delm}\bar C\right)\, \bar \del \del_-^2\, \left(\frac{1}{\delm} C\, \frac{1}{\delm} C\right)\ \nn \\ [0.2cm]
&&+\ 4\, \kappa^2\, \frac{\bar \del}{\del_-^2}\left(\del_-^3 \bar C\, \frac{1}{\delm}C\right)\, \frac{\del}{\del_-^2}\, \left(\del_-^3 C\, \frac{1}{\delm}\bar C\right)\  \\ [0.2cm]
&& + \ 4\, \kappa^2\Bigg\{2\, \del_-^2 \bar C \, \frac{\del}{\del_-^3}  \left(\del_-^3 C\, \frac{1}{\delm}\bar C\right)\, \frac{\bar \del}{\delm}C -\, \del_-^2 \bar C\, C\, \frac{\del \bar \del}{\del_-^4}\left(\del_-^3 C\, \frac{1}{\delm}\bar C\right)\, +\, c.c. \Bigg\} \nn \ .
\eea

\vskip 0.2cm
\ndt Here $\mathcal H^{\kappa^2}$ on the right-hand side stands for the $\kappa^2$-interaction terms obtained from the old Hamiltonian (\ref{H4}) with $h$ replaced by $C$ at the lowest order. Before we proceed to discuss the symmetries of the new Hamiltonian, we first study the role the field redefinition (\ref{FRD}) plays in the Hamiltonian formulation.
\vfill

\section{The field redefinition as a canonical transformation}

\ndt   In this section, we demonstrate how the non-local field redefinition can be viewed as an infinitesimal canonical transformation in the phase space. Henceforth we treat the Hamiltonian (\ref{H4}) as the fundamental entity and the Lagrangian is always derived from it using a Legendre transform\footnote{From our knowledge of $\mathcal L$ (\ref{d=4L}), we can define the conjugate momenta as
	\be
	\pi_h~=~ \frac{\delta \mathcal L}{\delta(\del_+ h)} ~=~  \del_- \bar h \ .
	\ee  But as we are working in the Hamiltonian formulation, we treat the conjugate momenta as independent variables and do not need to identify $\pi_h$ with $\delm \bar h$ or even define it in terms of $\mathcal L$. }
\bea
\mathcal{L}&=& \pi_h\, \del_+ h\ + \pi_{\bar h}\, \del_+ \bar h\ -\ \mathcal H \ ,
\eea

\ndt $\pi_h$ and $\pi_{\bar h}$ being the momenta conjugate $h$ and $\bar h$ respectively. The phase space variables $(h, \bar h, \pi_h, \pi_{\bar h})$ satisfy the following Poisson bracket relations

\bea
&& \{ h(x), \bar h(y) \} ~=~ \{ h(x), h(y) \}~=~ \{ \bar h(x), \bar h(y) \}= 0 \ , \\ [0.1cm]
&& \{ \pi_h(x), \pi_{\bar h}(y) \}~=~ \{ \pi_h(x), \pi_h(y) \}~=~ \{ \pi_{\bar h}(x), \pi_{\bar h}(y) \} ~=~ 0 \ ,\\ [0.1cm]
&& \{ h(x), \pi_h(y) \} ~=~ \{ \bar h(x), \pi_{\bar h}(y) \} ~=~ \delta^{(3)}(x-y)\ ,
\eea

\vskip 0.2cm
\ndt
where the Poisson brackets with $h, \pi_h$ are defined as
 \[\{A,B \}_{h, \pi_h} = \left(\frac{\delta A}{\delta h }\, \frac{\delta B}{\delta \pi_h}-\frac{\delta B}{\delta h}\, \frac{\delta A}{\delta \pi_h} \right)\] 

\ndt
and similarly with $\bar h, \pi_{\bar h}$. 

\vskip 0.2cm
To construct the Lagrangian corresponding to the ``new" Hamiltonian (\ref{H4new})
\bea \label{Lnew}
\mathcal{L'}&=& \pi_C\, \del_+ C\ + \pi_{\bar C}\, \del_+ \bar C\ -\ \mathcal H'\ ,
\eea

\vskip 0.2cm
\ndt we need to suitably define the conjugate momenta $\pi_C$ and $\pi_{\bar C}$ in order that the new fields satisfy the correct Poisson bracket relations as before, for e.g.,

\bea
\{ C(x), \pi_C(y) \} ~=~ \{ \bar C(x), \pi_{\bar C}(y) \} ~=~ \delta^{(3)}(x-y) \ .
\eea

\vskip 0.2cm

\ndt This is guaranteed if we can show that the field redefinition (\ref{FRD}) from $(h, \bar h)$ to $(C, \bar C)$ is  just an infinitesimal canonical transformation in the phase space leaving the Poisson brackets invariant, i.e.,

\bea
\{ h(x), \pi_h(y) \}_{h, \pi_h} ~=~ \{ C(x), \pi_C(y) \} _{h, \pi_h}~=~ \delta^{(3)}(x-y)\ .
\eea

\vskip 0.3cm
\subsection{The generating functional}

\ndt
We construct a generating functional of type 3 involving the new fields $C,\bar C$ and the old momenta $\pi_h, \pi_{\bar h}$ following the discussion in~\cite{Ananth:2007zy} 

\bea
G(C, \bar C, \pi_h, \pi_{\bar h})&=& \int\, g(C, \bar C) \, \pi_h	\ +\ \int\, \bar g(C, \bar C)\, \pi_{\bar h} \ ,
\eea
\vskip 0.2cm
\ndt
where we choose the function $g(C, \bar C)$ to be

\bea \label{gen func}
g(C) = C\ -\ \kappa \frac{\bar \del}{\del}\, \del_-^2\, \left(\frac{1}{\delm}C\, \frac{1}{\delm} C\right)\ -\ 2\, \kappa\, \frac{\del}{\bar \del}\, \frac{1}{\del_-^2}\, \left(\del_-^3 C\, \frac{1}{\delm} \bar C \right)\ +\ \mathcal{O}(\kappa^2)\ .
\eea

\vskip 0.2cm
We can derive the rest of the variables, namely the old fields and new momenta, from the generating functional $G$. The old field variables $h, \bar h$  can be derived from the generating functional as

\bea
h~=~ \frac{\delta G}{\delta \pi_h} ~=~ g(C, \bar{C})\ , && \bar h~=~ \frac{\delta G}{\delta \pi_{\bar h}}~=~ \bar g(C, \bar C)\ ,
\eea

\vskip 0.2cm
\ndt which reproduce our desired field redefinition (\ref{FRD}) as a canonical transformation. The new conjugate momentum $\pi_C$ is defined as 

\bea
\pi_{C}~=~ \frac{\delta G}{\delta C}~=~ \int \, \frac{\delta g}{\delta C}\, \pi_h~=~ \int \, d^3 x\  \pi_h (x)\ \frac{\delta h(x)}{\delta C(y)}\ ,
\eea

\vskip 0.2cm
\ndt where we have used the fact that $h=g(C, \bar C)$. Similarly $\pi_{\bar{C}}$ can be calculated from (\ref{gen func}) as the functional derivative of $G$ with respect to $\bar C$. 

\vskip 0.2cm
We can now check if the canonical transformation generated by the functional $G$ preserves the Poisson bracket relations.

\bea
\{ C (x), \pi_{C}(y) \}_{h, \pi_h}&=&	\left( \frac{\delta C(x)}{\delta h(z)}\,\frac{\delta \pi_C(y)}{\delta \pi_h(z)}- \frac{\delta C(x)}{\delta \pi_h(z)}\, \frac{\delta \pi_C(y)}{\delta h(z)} \right) \nn \\ [0.2cm]
&=& \frac{\delta C(x)}{\delta h(z)}\, \frac{\delta}{\delta \pi_h(z)}\left\{\, \int \, d^3 y'\  \pi_h (y')\ \frac{\delta h(y')}{\delta C(y)}\, \right\} \ - \ 0 \nn \\ [0.2cm]
&=& \int \, d^3 y'\  \frac{\delta C(x)}{\delta h(z)} \, \delta^{(3)}(z-y')\, \frac{\delta h(y')}{\delta C(y)} \nn \\ [0.2cm]
&=& \frac{\delta C(x)}{\delta h(z)} \, \frac{\delta h(z)}{\delta C(y)}~=~ \delta^{(3)}(x-y) \ .
\eea

\vskip 0.2cm	
\ndt Therefore, this canonical transformation maps the original light-cone Hamiltonian to the new Hamiltonian without any cubic vertices, which now describes the theory in terms of the fields $(C, \bar C)$.

\subsection{The field redefinition as a change of variables}
\ndt We are now in a position to define an action functional in the Hamiltonian formulation as

\bea
\mathcal S(h, \bar h) &=& \int d^4 x \left(\pi_h\, \del_+ h\ + \pi_{\bar h}\, \del_+ \bar h\ -\ \mathcal H\right)~=~ \int\, d^4x\, \mathcal{L}(h, \bar h)
\eea

\ndt with the corresponding path integral given by

\bea \label{PI}
\mathcal{I}&=& \int \, [\mathcal Dh]\ [\mathcal D \bar h] \ e^{i\mathcal S(h, \bar h)}\ .
\eea

\ndt
We can similarly define an action with the the new fields 

\bea 
\mathcal S(C, \bar C)&=& \int\, d^4 x\, \mathcal{L'}(C,\bar C) \ .
\eea

\vskip 0.2cm
We can argue that the field redefinition amounts to a change of variables in the path integral with the action $S(h, \bar h)$ replaced by $S'(C, \bar C)$, if the integration measure $[\mathcal D...]$ remains invariant under the transformation. Under a general transformation, the integration measure changes as

\bea
\int \, [\mathcal Dh]\ [\mathcal D \bar h] &\rightarrow& \int \,(det\ J)  \ [\mathcal D C]\ [\mathcal D \bar C]\ ,
\eea

\ndt where $J$ stands for the Jacobian of the transformation. 

\vskip 0.2cm
In our perturbative approach, the field redefinition is an expansion in orders of $\kappa$, i.e.,  $h= C\ + \kappa (\cdots)$. Hence, it easily follows that $det\ J=1 +\ \mathcal O(\kappa)$ in our case. Therefore, at the lowest order the path integral in terms of $C, \bar C$ reads

\bea
\mathcal{I'}&=& \int \, [\mathcal D C]\ [\mathcal D \bar C] \ e^{i\mathcal S(C, \bar C)}\ ,
\eea

\ndt which is simply a change of variables in (\ref{PI}). This essentially means both the path integrals $\mathcal{I}$ and $\mathcal{I'}$ describe the same theory and will give rise to the same correlation functions or scattering amplitudes. 

\vskip 0.2cm
The crux of the problem is that the helicity states $(h, \bar h)$ in the original action are not the ``good" field variables for representing the enhanced symmetry. We must make a change of variables to the appropriate fields $(C, \bar C)$ to make the symmetry manifest. Having shown that the two formulations describe the same theory, we can now proceed to investigate the symmetries the new Hamiltonian exhibits, which are obscure in the $h, \bar h$ configuration.

\vskip 0.2cm

\section{Ehlers symmetry in $d=4$}

\ndt To find the Ehlers symmetry in four dimensions, we start with a sigma-model like transformation with parameters $a$ and $\bar a$

\bea \label{symm}
\delta C = \frac{1}{\kappa}\, a\ , && \delta {\bar C}~=~ \frac{1}{\kappa}\, \bar a\ 
\eea

\vskip 0.2cm

\ndt and consider the variation of the new Hamiltonian (\ref{H4new}) under these transformations

\bea
\mathcal{H'}&=& \mathcal{H'}^{0}\ +\ \mathcal{H'}^{\kappa^2}~=~\bar \del C\, \del \bar C\ +\ \kappa^2\, (\cdots)\ .
\eea

\vskip 0.2cm
\ndt At the lowest order, the Hamiltonian varies as

 \bea
(\delta  \mathcal H')^{\kappa^{-1}}&=& \bar \del (\delta C)\, \del \bar C\ + \bar \del C\, \del (\delta \bar C)\ , \nn
 \eea
 which vanishes trivially since the parameters $a$ and $\bar a$ are constants. At the next order, we have
 \bea \nn
 (\delta \mathcal H')^\kappa &=& \delta^{\kappa}\, \mathcal H'^0\ +\ \delta^{\kappa^{-1}}\, \mathcal H^{\kappa^2} \ .
 \eea
 
 \ndt
The variation of the interacting part of the Hamiltonian, $\delta^{\kappa^{-1}}\, \mathcal H^{\kappa^2}$ can be compensated for by adding correction terms of order $\kappa$ terms to (\ref{symm})

\bea \label{fields}
\delta C &=& \frac{1}{\kappa} a\ + \ 2\, \kappa\, a\, \frac{1}{\delm}(\delm C \,\bar C)\ -\ \kappa\, \bar a\, C\, C \ ,\nn \\
\delta\bar C&=& \frac{1}{\kappa} \bar a\ + \ 2\, \kappa\, \bar a\, \frac{1}{\delm}(\delm \bar C \,C)\ -\ \kappa\, a\, \bar C\ \bar C \ ,
\eea

\ndt which act on $\mathcal H'^0$, thereby rendering the Hamiltonian invariant up to order $\kappa^2$. Therefore, the transformations (\ref{fields}) correspond to a symmetry of the new Hamiltonian. 

\vskip 0.2cm 
To reveal the Lie algebra underlying this symmetry, we rewrite these transformations as a set of two transformations parameterized by $a$ and $\bar a$ respectively

\bea
L_+\, C&=& \frac{1}{\kappa} a\ + \ 2\, \kappa\, a\, \frac{1}{\delm}(\delm C \,\bar C)\ , \nn \\ L_+\, \bar C&=& \ -\ \kappa\, a\, \bar C\, \bar C\ 
\eea
\bea
 L_-\, C&=&\ -\ \kappa\, \bar a\, C\, C\ , \nn \\L_-\, \bar C&=& \frac{1}{\kappa} \bar a\ + \ 2\, \kappa\, \bar a\, \frac{1}{\delm}(\delm \bar C \,C)\ .
\eea

\vskip 0.2cm
\ndt We augment the above transformations with an $SO(2)$ transformation acting linearly on the fields  
\bea
L_0 \, C~=~ 2\, \bar a\, a\, C\ , && L_0 \bar C~=~ -\, 2\, \bar a\, a \bar C\ .
\eea

\vskip 0.2cm
\ndt We find that these $L_+, L_-$ and $L_0$ transformations satisfy an SL(2,R) algebra
\bea
[L_+, L_-]\ = \ L_0 \ , && [L_0, L_\pm]\ =\ L_\pm \ .
\eea

 \vskip 0.2cm
\ndt Thus, we recover the Ehlers symmetry in $d=4$ light-cone gravity found in~\cite{Ananth:2018hif}, without having to reduce the theory to three dimensions. We must remember though that the symmetry is only strictly proven to order $\kappa^2$, but we do not expect any formal difficulties in extending these results to higher orders. However, the explicit calculations become quite intractable to be done analytically since the number of terms at each order increases tremendously.

\vskip 0.2cm
The $L_0$ transformations seem to have been added ad hoc for the algebra to close. These transformations mimic the action of the little group rotations in four dimensions, which assigns helicity values to the fields. On the contrary, this $SO(2)$ is inherently different from the helicity group as we discuss below.

\subsection {$SO(2)$ little group versus duality symmetry}
\ndt Little group transformations in four  dimensions, which rotate the transverse coordinates and derivatives into each other, are generated by the  $J^{12}$ element of the light-cone Poincar\'e algebra. In the massless case, its spin part $S^{12}$ determines the helicity of the fields~\cite{Bengtsson:1984rv}, for e.g.,
\bea \label{helicity}
S^{12}\, \left(\begin{array}{c}
	h \\ \bar h
\end{array}\right) &=& 2\, \left(\begin{array}{c}
	h \\ -\bar h
\end{array}\right)\ .
\eea

\vskip 0.2cm
\ndt The light-cone Lagrangian (\ref{d=4L}) for gravity expressed in terms of these fields is thus helicity-invariant, even though Lorentz invariance is no longer manifest. In particular, the cubic interaction vertices are manifestly helicity-invariant because the derivatives $\del$ and $\bar \del$ also transform under the SO(2). 

\vskip 0.2cm 
We now return to the $L_0$ transformations, which form the maximal compact subgroup of the $SL(2, \R)$ discussed in the last section
\bea \label{SO(2)}
L_0\, \left(\begin{array}{c}
	C \\ \bar C
\end{array}\right) &=& 2\, \bar a a\, \left(\begin{array}{c}
C \\ -\bar C
\end{array}\right)\ .
\eea
 
 \vskip 0.2cm
 \ndt It seems that $L_0$ can be identified with the helicity group (\ref{helicity}), if we set $\bar a\, a =1$. But as the $SL(2, \R)$ is a hidden symmetry of the theory, the SO(2) subgroup must also be a part of this internal symmetry and not the Poincar\'e group.

\vskip 0.2cm
This $SO(2)$ is a remnant of the electro-magnetic duality symmetry in gravity, which leaves the equations of motion invariant under the exchange of the electric and magnetic part of the curvature tensor~\cite{Henneaux:2004jw}. In the light-cone formalism, since we have fixed the gauge completely and eliminated all the unphysical degrees of freedom, the equations of motion are not expressed in terms of the curvature, but the fields $C$ and $\bar C$. Therefore, the electro-magnetic duality symmetry is also translated in terms of rotations in the two remaining degrees of freedom of the graviton. The new fields $C$ and $\bar C$ are, thus, the eigenstates of the duality operator $L_0$, just as the old fields $h$ and $\bar h$ are the eigenstates of the helicity operator $S^{12}$.

\vskip 0.2cm
Unlike the little group transformations, the SO(2) duality does not act on the derivatives $\del$ and $\bar \del$. As a result, it is not a symmetry of the original light-cone action due to the presence of cubic vertices. To realize this symmetry, we must give up the {\it{helicity-invariant}} formulation in favor of the {\it{duality-invariant}} one, where we have removed the cubic vertices through a suitable field redefinition. This interpretation resonates with the idea that hidden symmetries in gravity and supergravity theories arise from a trade-off between the spacetime symmetries and the internal symmetries.

\vskip 0.2cm
In our approach the fields depend on all four spacetime coordinates in both the formulations. In spirit, this is similar to the dual graviton and exceptional field theories~\cite{Hohm:2013jma, Hohm:2013pua}, where one considers a $D=n+d$ split in a $D$-dimensional theory with certain constraints to be fulfilled by the $d$-dimensional internal space. The key difference is that their approach is manifestly covariant and local, while ours is not. Nonetheless, the crucial point is that we are not performing any dimensional reduction to uncover the enhanced symmetry in $d=4$, neither are we constraining the dynamics of the theory to a three-dimensional subspace. 

\section{Concluding remarks}

\vskip 0.2cm
\ndt In the light-cone superspace, $\mathcal{N}=8$ supergravity in four dimensions can be shown to exhibit an $E_{8(8)}$ symmetry, enhanced from the previously known $E_{7(7)}$ symmetry~\cite{Ananth:2017nbi}. In order to realize this symmetry enhancement, one must abandon the helicity states, originally in representations of the $SU(8)$ R-symmetry group, in favor of an $SO(16)$ representation which mixes fields of different spins. In~\cite{Ananth:2018hif}, the Ehlers symmetry in four dimensions was obtained as a remnant of this $E_8$ symmetry in $\mathcal N=8$ supergravity after supersymmetric truncation to pure gravity. The obvious next step would be to repeat the analysis presented here for the $\mathcal{N}=8$ theory to find a suitable map from the helicity states to the appropriate field variables, on which the $E_{8(8)}$ symmetry can be made manifest in four dimensions. 
\vskip 0.2cm
The important lesson we learn is that only in the right field configuration can we see the various symmetries the theory possesses. In order to find all the symmetries, we have to know the correct field redefinitions that allow us to shift between various configurations. This opens the door to many intriguing possibilities. One of the most pressing questions it raises is whether the Ehlers symmetry is the only hidden symmetry one can realize in four-dimensional Einstein's gravity or is there room for more. The immediate candidate for a further symmetry enhancement would be the Geroch group in Einstein's theory. On reduction to two dimensions, the theory has an $SL(2,\R)$ symmetry called the Matzner-Misner group~\cite{Matzner:1967zz}, which does not commute with the existing Ehlers $SL(2, \R)$ group; instead they generate the infinite-dimensional Geroch group. To understand the Geroch group in the light-cone language, first we must find the right formulation to realize the Matzner-Misner symmetry. For supergravity theories, one can pose similar questions: Is $E_{8(8)}$ symmetry the largest symmetry group one can realize in four dimensions or can we extend it further to $E_9$ and $E_{10}$? Can we find a field redefinition in eleven-dimensional supergravity, that maps the theory to a different configuration with manifest $E_{7(7)}$ or $E_{8(8)}$ symmetry? 

\vskip 0.2cm

It has been established time and again that the Hamiltonian formulation is better suited than the Lagrangian formulation for the study of hidden symmetries off-shell. In~\cite{Henneaux:2004jw}, the duality-invariant action for linearzied Einstein's gravity was constructed using prepotentials in the Hamiltonian formulation. We obtain a duality-invariant formulation for light-cone gravity without restricting the theory to the linearized order. Admittedly, our perturbative results are not sufficient to comment on the nature of the full non-linear theory. Nevertheless, the fact that there exists a formulation for gravity in four dimensions with manifest Ehlers symmetry suggests that a non-linear extension of the duality-invariant action in~\cite{Henneaux:2004jw} might lie within reach.

\vskip 0.2cm

The prepotential formalism has been instrumental in constructing several first-order Hamiltonian actions for self-dual fields, which include the free actions for $\mathcal N= (4,0)$ and $\mathcal N=(3,1)$ supergravity in six dimensions~\cite{Henneaux:2017xsb,Henneaux:2018rub} to name a few. But the construction of interacting actions for these self-dual fields has been a longstanding problem, for there exists a plethora of no-go theorems that rule out such interactions on the grounds of Lorentz covariance and locality (see~\cite{Bekaert:2000qx,Bekaert:2002uh,Bekaert:2004dz,Boulanger:2004rx} for a non-exhaustive list). The light-cone formalism, on the contrary, violates manifest Lorentz covariance and allows for a unique type of non-locality in the spatial coordinates. These key ingredients seem to do the trick for constructing a duality-invariant formulation for gravity at least up to the second order in coupling constant. In light of these results, it would be interesting to see if the inclusion of certain elements of non-locality in the prepotential formalism can help circumvent the aforementioned no-go results. Likewise, one could try to incorporate similar tools into the powerful framework of BRST cohomology for gauge theories~\cite{Barnich:1993vg,Barnich:2000zw}, in order to explore the notion of spatial non-locality with greater mathematical rigor. It would be nice to further extend the light-cone analysis of gravity to establish links to other interesting symmetry-related directions~\cite{Strominger:2013jfa}.

\section{Acknowledgments}

\ndt We thank Sudarshan Ananth and Lars Brink for insightful discussions and for helpful comments on the manuscript. We are grateful to Glenn Barnich, Marc Henneaux and Stefan Prohazka for many stimulating discussions. This work was partially supported by a Marina Solvay Fellowship, by the ERC Advanced Grant ``High-Spin-Grav" and by FNRS-Belgium (convention FRFC PDR T.1025.14 and convention IISN 4.4503.15). We acknowledge the kind hospitality offered during the workshop ``Higher spins and Holography" at the Erwin Schr\"odinger Institute in Vienna, where a part of this work was completed.
	\appendix
	\section{Gravity in the light-cone gauge}
	
	\ndt In this appendix we briefly discuss the light-cone gauge-fixing and perturbative expansion of the Einstein-Hilbert action in four dimensions.

	\vskip 0.2cm
	The Einstein-Hilbert action on a Minkowski background reads 
	
	\bea
	S_{EH}=\int\,{d^4}x\;\mathcal{L}\,=\,\frac{1}{2\,\kappa^2}\,\int\,{d^4}x\;{\sqrt {-g}}\,\,{\mathcal R}\ ,
	\eea
	\vskip 0.1cm
	\ndt
	with the corresponding field equation
	
	\bea
	\label{feq}
	\mathcal R_{\mu\nu} \,-\,\half\, g_{\mu\nu}\mathcal R\,=\,0\ .
	\eea
	\vskip 0.2cm
	\ndt where the symbols have their usual meaning. We now impose {\it {three}} of the four allowed gauge choices~\cite{Scherk:1974zm,Bengtsson:1984rv}
	\bea \label{lcg}
	g_{--}\,=\,g_{-i}\,=\,0\quad ,\; i=1,2\ .
	\eea
	
	\ndt The rest of the  metric components are parametrized as 
	\bea
	\label{gc}
	\begin{split}
		g_{+-}\,&=\,-\,e^\phi\ , \\
		g_{i\,j}\,&=\,e^\psi\,\gamma_{ij}\ .
	\end{split}
	\eea
	\vskip 0.2cm
	\ndt $\phi, \psi$ are real parameters and $\gamma^{ij}$ is a real, symmetric matrix with unit determinant. As a result of the light-cone gauge choice, some of the field equations do not involve time derivatives $(\del_+)$ and become constraint relations which can be solved to eliminate more degrees of freedom from the theory. The $\mu\!=\!\nu\!=\!-\;\;$equation from (\ref {feq}) is such a constraint relation which gives 
	\bea \label{CE1}
	2\,\del_-\phi\,\del_-\psi\,-\,2\,\del^2_-\psi\,-\,(\del_-\psi)^2\,+\, \half \del_-\gamma^{ij}\,\del_-\gamma_{ij}\,=\,0\ .
	\eea
	\vskip 0.2cm
	We now make our {\it {fourth}} gauge choice 
	\bea 
	\phi\,=\,\frac{\psi}{2}\ ,
	\eea 
	which allows us to solve for $\psi$ in terms of $\gamma_{ij}$. Similarly we can use the other constraint relations to eliminate all the unphysical degrees of freedom and obtain the light-cone action for gravity in a closed form expression
	
	\bea
	\label{aaction}
	S\,&=&\frac{1}{2\kappa^2}\int d^{4}x \; e^{\psi}\left(2\,\del_{+}\del_{-}\phi\, +\, \del_+\del_-\psi - \half\,\del_{+}\gamma^{ij}\del_{-}\gamma_{ij}\right) \nonumber \\
	&&-e^{\phi}\gamma^{ij}\left(\del_{i}\del_{j}\phi + \half \del_{i}\phi\del_{j}\phi - \del_{i}\phi\del_{j}\psi - \frac{1}{4}\del_{i}\gamma^{kl}\del_{j}\gamma_{kl} + \half \del_{i}\gamma^{kl}\del_{k}\gamma_{jl}\right) \nn \\
	&&- \half e^{\phi - 2\psi}\gamma^{ij}\frac{1}{\del_{-}}R_{i}\frac{1}{\del_{-}}R_{j}\ ,
	\eea
	\ndt where
	 
	\bea 
	R_{i}\,\equiv\, e^{\psi}\left(\half \del_-\gamma^{jk}\del_{i}\gamma_{jk}-\del_-\del_i\phi - \del_-\del_i\psi + \del_i\phi\del_-\psi\right)+\del_k(e^\psi\,\gamma^{jk}\del_-\gamma_{ij})\ . \nn
	\eea
	
	\ndt
	\subsubsection*{Perturbative expansion}
	\vskip .3cm
	\ndt We consider a perturbative expansion of the closed form action where we parameterize $\gamma_{ij}$ as~\cite{Bengtsson:1984rv}
	\bea 
	\gamma_{ij}\,=\,(e^{\kappa\, H})_{ij}\ ,
	\eea
	\vskip 0.2cm
	\ndt
	where $H$ is a traceless matrix since $det\,(\,\gamma_{ij})=1$. We choose
		\bea \label{matrixH}
	H\,=\,\begin{pmatrix} h_{11} & h_{12} \\ h_{12} & -h_{11} \end{pmatrix}\ ,
	\eea
	
	\vskip 0.2cm
	\ndt where $h_{11}$ and $h_{12}$ can be linearly combined to form the helicity states of the graviton
	
	\bea
	 h\,=\,\frac{(h_{11}+i\,h_{12})}{\sqrt{2}}\,, && \bar{h}\,=\,\frac{(h_{11}-i\,h_{12})}{\sqrt{2}}\ .
	\eea

\vskip 0.2cm
	The light-cone Lagrangian can then be expressed perturbatively in orders of the coupling constant $\kappa$. To the first order in $\kappa$, it reads
	
	\bea 
	\label{kinetic}
	\mathcal{L}_2\,=\, \half \,\bar h\,\Box\, h\ \ +\ \, 2\,\kappa\, \bar{h}\, \del_-^2\left[-\,h\,\frac{\bar{\del}^2}{\del_-^2}h\,+\,\frac{\bar{\del}}{\del_-}h\,\frac{\bar{\del}}{\del_-}h\right] \,+\, \text{c.c.}\ ,
	\eea
	
	\vskip 0.2cm
	\ndt with the light-cone d'Alembertian defined as
	\bea
	\Box\,=\,2\,(\,\partial\,{\bar \partial}\,-\,\partial_+\,\del_-\,)\ .
	\eea
	\vskip 0.2cm
	\ndt  The Lagrangian to order $\kappa^2$ was presented in~\cite{Bengtsson:1984rv,Ananth:2007zy}, while the order $\kappa^3$ corrections can be found in~\cite{Ananth:2008ik}.

\end{document}